# UPGRADE OF THE CONTROL SYSTEM OF THE IFUNAM´S PELLETRON ACCELERATOR

R. Macías, E. Chávez, M. E. Ortiz, K. López, A. Huerta, IFUNAM, A. P. 20-364, México 01000
M. C. Verde, Instituto de Ingeniería UNAM, México

Abstract

In 1995 a 9SDH-2 Pelletron from NEC was installed at IFUNAM (Instituto de Física, Universidad Nacional Autónoma de México). Two beam lines have been operational since then and two new lines have been built. In order to perform the planned projects in this grown facility, an upgrading of the original manual control system is required. The proposed new control system takes advantage of the existing devices and incorporates the electronics needed for the newer beam lines. The control software from NEC, has been modified to accommodate the larger requirements. It runs on the same dedicated computer but receives commands from a new installed host. Both computers communicate through a local network sharing the accelerator database. The new host computer also handles all parameters related to the new lines. In the future, the old computer will be replaced in order to expand the possibilities of the system and use a friendlier graphical interface. In this work we present the changes made to the control software, the new acquisition and control devices installed and the integration of the old and new systems. Finally, the beam control procedure based on the analysis of the beam profile that is under development is also shown.

## INTRODUCTION

After arrival and installation of the IFUNAM's pelletron, it was soon recognized the need for additional beam lines. The first two, those that originally came with the machine, are designed to be dedicated to Rutherford Back Scattering (RBS) Analysis and Ion Implantation respectively. Two new lines were proposed, one to be dedicated to Particle Induced X-Ray Analysis (PIXE) and a second for Nuclear Reaction Studies and Mass Spectrometry.

The original computer control system must be expanded to be able to handle the new hardware added to the system. Two options were envisaged: build a whole new control system or modify the existing one. The former requires more resources both in time, money, skilled personal and deep knowledge of the manufacturer's design. The last one, less ambitious, takes advantage of the existing software and hardware and is easier to implement. Additionally, from the standpoint of the operator, most of the operations remain unchanged. We decided to take this line of work, at least in a first stage, keeping in mind that a new system could be built in the future.

## SYSTEM OVERVIEW

The control system [1] acts primarily as a switchyard which takes incoming signals from the accelerator and displays the parameters on the screen and or meters. The operator initiates control of the various parameters. The control system software, developed using MS Visual Basic for DOS, includes: a) The control system program to interact with the operator and database to provide control and monitoring of the accelerator and b) a number of utilities programs.

Each component of the accelerator has a unique label and reference number, that may be associated with a process control station address, a special control/read, a bypass or an interlock record. Each record has two separate files or database. One file is for data that does not change (cs1.db) and the other is for items that change (cs.db). These records hold the information on address, slot, size, type, etc. for the control system program to control and/or monitor all data points in the system.

## THE MODIFICATIONS

A second computer was added (PIII, 128MB RAM, hereupon referred as S2), partly because the original program only runs properly in its host computer (Digital 486, 4MB RAM, we will call this S1 from now on). Both computers were connected through an Ethernet link in such a way that S2 could access the working directory of S1.

The original program from NEC on S1 was modified in order to allow the program running on S2 to remotely interact with the database in S1.

The program on S1 remains the only one with direct access to the database. The interaction between S1 and S2 programs is made through two new files (updrec.db and readrec.db), so S2 can request the modification or readout of the value of any system parameter. Two new procedures were added to the program in S1, to force it to act according to the requests from S2.

In this way the system can be controlled from an external computer, which allows an easy way to expand the original control system, adding additional software in the external computer without any other modification of the original one.

As an example the Fig. 1 shows the front panel of a procedure (written in LabVIEW 5.1 from National Instruments Corp.) for updating any parameter of accelerator. The control rec number identifies the accelerator parameter (i.e. 78 means Injector Magnet Current) and pdatareal is the real value we want to update.

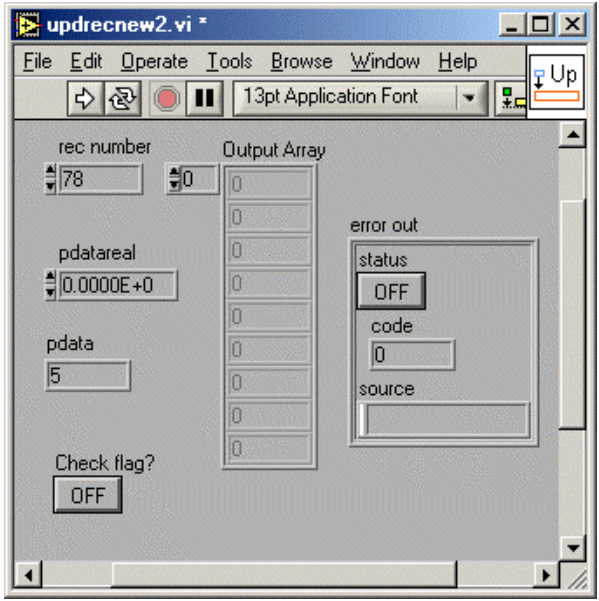

Figure 1: Front panel of the updating procedure.

## ONE ADDITIONAL BEAM LINE

Here we describe one of the new beam lines added to the accelerator: the isotope separator.

The 45° port of the switching magnet was selected for it. The separator consists mainly of electromagnetic devices for the transport of the beam, together with beam control and diagnostics as shown schematically in Fig. 2.

These additional elements are controlled by S2 through FieldPoint modules from National Instruments Corp., connected to RS232 port by optical fiber.

Besides all the components of the new line, three teslameters model DTM-141 from Group3 Technology Ltd., were added to the accelerator system to have a precise measurement of the magnetic induction in the injector, switching and 90° magnets.

Further expansion of the laboratory: new beam lines or additions to the existing ones, can be computer controlled following the procedures here described.

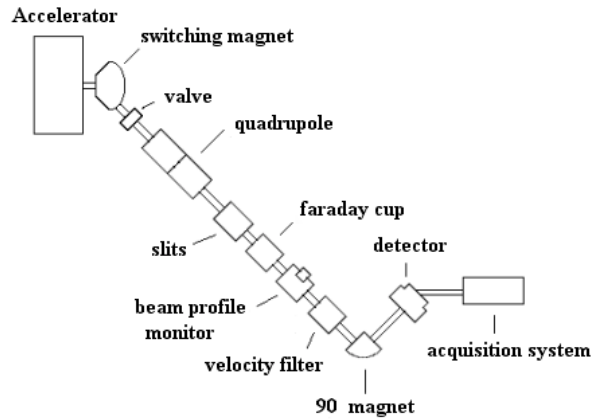

Figure 2: Isotope Separator on the 45º port.

## BEAM CONTROL AND DIAGNOSIS

The accelerator is equipped with beam profile monitors model BPM-80 from NEC at the low energy side and in all of the beam lines. This kind of device allows continuous monitoring of the intensity distribution and the position of the beam [2].

Its working principle is the collection of secondary electrons released from a rotating wire that sweeps across the beam twice in each cycle, at a constant speed. The resulting signal (X and Y profiles) is amplified and sent to the control console, together with fiducial marks. These marks are time signals generated internally by the BPM. All these signals are displayed on an oscilloscope (see Fig. 3).

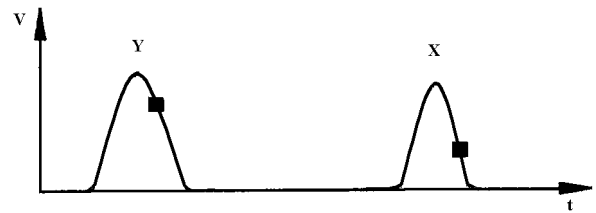

Figure 3: Profile with marks as observed on the control console oscilloscope.

The machine operator uses the scope image to obtain the desired characteristics of the requested beam (focusing and position) by looking at the width, height and location of the peaks relative to the fiducials.

Commonly the intensity of the beam is measured through the use of a Faraday Cup (FC). This requires to stop the beam. A new procedure is developed here to use the BPM signals to obtain information about the beam intensity, with the advantage of not having to interrupt it.

An electronic shaping circuit has been designed. This circuit will allow the acquisition of signals from two

BPMs by the S2 computer, equipped with a PCI-6024E card from National Instruments Corp.

Using a 3 MeV proton beam, 135 profiles of the kind of Fig. 3 were recorded for different combinations of focus, cathode, extraction and bias voltages. These are all the relevant parameters of the SNICS ion source. For each record the beam current (Ifc) was measured with a FC.

Fig. 4 shows a plot of the measured current (Ifc) and a value extracted by the product of the integrals of the two peaks in the profile (Int(x) * Int(y) * N). The normalization factor N depends on the beam species, energy, amplifiers gain, etc.

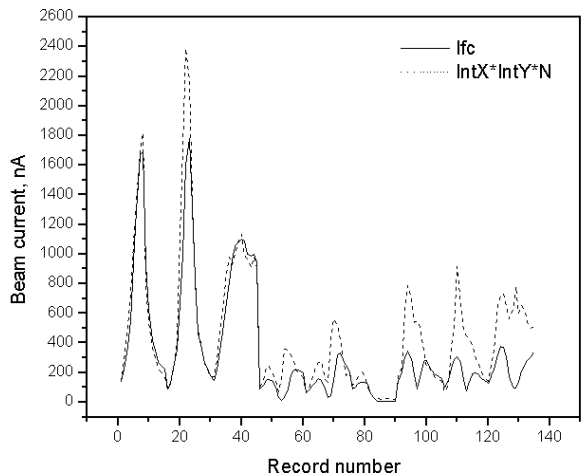

Figure 4: Comparison between the measured and calculated beam currents.

One can notice that both curves follow each other closely up to record 45. From this point an important deviation appears, which can be explained in terms of a displacement of the beam off the center of the geometrical axes that were no taken into account during calculations.

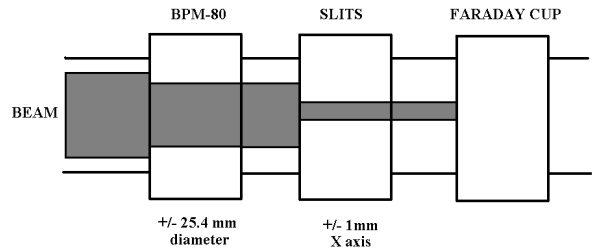

Figure 5: BPM, slits and Faraday Cup as located in the low energy side of accelerator.

The integration limits needed above are determined mainly by the aperture of the slits and BPM as shown in Fig. 5.

Data from the first 45 records is shown in Fig. 6, where the calculated beam current from the BPM is plotted versus the Ifc. The straight line shows the best fit to the data.

This behavior allows us to conclude that the beam current can be measured using the BPM, avoiding in this way the interruption of the beam and permitting continuous and controlled operation of the experiment.

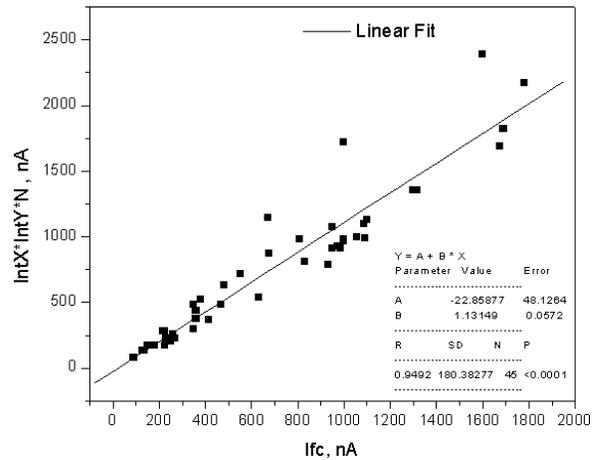

Figure 6: Correlation between the measured and calculated beam currents.

It is worth mentioning that the position of the fiducial marks on each profile is critical and, in order to get better results, they should be accurately measured for every cycle.

## ACKNOWLEDGEMENTS

The authors wish to thank the IFUNAM's Pelletron support personnel, specially Francisco X. Jaimes, without whom this work would not have been possible. This work was supported by UNAM through contracts DGAPA IN114896 and CONACyT 32262-E